\newif\ifsubmode
\submodefalse

\newif\ifprintfig
\printfigtrue

\ifsubmode
  \documentclass{aastex}
  \usepackage{epsfig}
  \received{}
  \accepted{}
  \journalid{}{}
  \articleid{}{}
\else
  \documentclass[preprint]{aastex}
  \usepackage{epsfig}
  \slugcomment{{\it STScI Newsletter, 2006, vol.~23, number 2, in press}}
\fi

\shorttitle{}
\shortauthors{}

\begin{document}

\title{Gravity's Relentless Pull: An interactive, multimedia website about
black holes for Education and Public Outreach}

\author{Roeland P.~van der Marel}
\affil{Space Telescope Science Institute, 3700 San Martin Drive, 
       Baltimore, MD 21218\\(marel@stsci.edu)}

\author{David Schaller}
\affil{Educational Web Adventures, 1776 Iglehart Avenue,
       St. Paul, MN 55104 \\(david@eduweb.com)}

\author{Gijs Verdoes Kleijn}
\affil{Kapteyn Astronomical Institute, Groningen, 9700 AV, The Netherlands
       \\(g.verdoes@astro.rug.nl)}



\ifsubmode\else
\clearpage\fi


\ifsubmode\else
\baselineskip=14pt
\fi


\begin{abstract}
We have created a website, called "Black Holes: Gravity's Relentless
Pull", which explains the physics and astronomy of black holes for a
general audience. The site emphasizes user participation and is rich
in animations and astronomical imagery. It won the top prize of the
2005 Pirelli INTERNETional Awards competition for the best
communication of science and technology using the internet. This
article provides a brief overview of the site. The site starts with an
opening animation that introduces the basic concept of a black
hole. The user is then invited to embark on a journey from a backyard
view of the night sky to a personal encounter with a singularity. This
journey proceeds through three modules, which allow the user to: find
black holes in the night sky; travel to a black hole in an animated
starship; and explore a black hole from up close. There are also five
``experiments'' that allow the user to: create a black hole; orbit
around a black hole; weigh a black hole; drop a clock into a black
hole; or fall into a black hole. The modules and experiments offer
goal-based scenarios tailored for novices and children. The site also
contains an encyclopedia of frequently asked questions and a detailed
glossary that are targeted more at experts and adults. The overall
result is a website where scientific knowledge, learning theory, and
fun converge. Despite its focus on black holes, the site also teaches
many other concepts of physics, astronomy and scientific thought. The
site aims to instill an appreciation for learning and an interest in
science, especially in the younger users. It can be used as an aid in
teaching introductory astronomy at the undergraduate level.
\end{abstract}



\clearpage


\section{Introduction}
\label{s:intro}

Some of the Hubble Space Telescope's most ground-braking discoveries
have been about black holes, which are arguably the most extreme and
mysterious objects in the universe. As a result, black holes appeal to
a general audience in a way that almost no other scientific subject
does. Unfortunately, most people have little idea of what black holes
actually are, and they are more likely to associate them with science
fiction than with science. These circumstances recommend black holes
as a natural topic for an education and public outreach (E/PO)
activity. To this end, we have created a website called "Black Holes:
Gravity's Relentless Pull," which serves as the E/PO component of
several Hubble observing projects. The new website is part of
HubbleSite, the Internet home of all Hubble news. The URL is
\underline{http:/$\!$/hubblesite.org/go/blackholes/} .

Many sites on the internet already explain black holes in one way or
other. Most are encyclopedic, offering detailed text and graphics.  By
contrast, our site emphasizes user participation, and it is rich in
animations and audio features. This approach was facilitated by the
availability of powerful software for authoring multimedia
content. The result is a website where scientific knowledge, learning
theory, advanced technology, and pure fun all converge.

\section{The Website}
\label{s:website}

The opening animation at our website introduces the basic concept of a
black hole. It shows how one could turn the Earth into a black hole,
if only one could shrink it to the size of a marble. By connecting
enigma and commonplace, this scenario lends black holes a seeming
familiarity.

The core of the website consists of three sequential, interactive
modules, of which "Finding the Invisible" is the first. It shows the
night sky with a viewfinder that can be dragged around to discover
images of about a dozen objects, including some that should be
familiar to the user (the Sun, Moon, and Saturn) and others that only
an astronomer might recognize (Betelgeuse, Crab nebula, Cygnus X-1,
Andromeda galaxy, and quasar 3C273). The user can select the
wavelength range of the viewfinder: visible light, radio waves, or
X-rays. The goal is to teach which types of objects contain black
holes and which do not. Interested users can move to pages that
explain the various objects, the telescopes used to observe them, and
the features in the images at different wavelengths that indicate the
presence of a black hole. In this way the user not only learns about
black holes, but also about their relation to other objects in the
Universe and the methods that astronomers use to study them.

When the user has found one or more black holes, he or she can choose
to go to the second module, "The Voyage," which offers a multimedia
trip in an animated starship to a nearby black hole, either a
stellar-mass black hole (Cygnus X-1) or a supermassive black hole (in
the center of the Andromeda galaxy). The viewer traverses the Solar
System as well as our own Milky Way galaxy. Various intriguing objects
are encountered along the way, including many of the objects
previously encountered in the first module. A goal is to connect the
two-dimensional, projected view of the night sky to the actual
three-dimensional structure of the universe. In the process, the user
learns about the distance scale and the layout of the local universe.

After the spaceship has arrived at the black hole, the user can
proceed to the third module, "Get Up Close." Orbiting around a black
hole, the sky outside the spaceship window looks strangely distorted,
because of the strong gravitational lensing of distant starlight. The
module discusses this and many other fascinating phenomena to be
encountered near black holes.

Five interactive experiments are offered to explore specific
issues. "Create a Black Hole" allows the user to study the evolution
of stars of different masses by trying to create a black hole, rather
than a white dwarf or neutron star. "Orbit around a Black Hole" plots
relativistically correct orbits around a black hole, showing how it is
possible to orbit a black hole without being sucked in. "Weigh a Black
Hole" shows how to use observations of a black hole in a binary system
to calculate its mass. This illustrates learning without seeing. "Drop
a Clock into a Black Hole" explains time dilation and
redshift. Mysteriously, an outside observer never sees a falling clock
disappear beyond the event horizon, which highlights the principle of
relativity. The last experiment, "Fall into a Black Hole" allows the
viewer to take the final plunge and witness how one's body gets
stretched by tidal forces. With this, the user's journey is complete,
from the backyard view of the night sky to a personal encounter with a
singularity.

\section{Learning theory}
\label{s:learning}

Our motivations for adopting a strongly interactive website were
grounded in specific learning theories, "constructivism" in
particular. Constructivism holds that learning is not merely an
addition of items into a mental data bank, but rather requires a
transformation of concepts in which the learner plays an active role
making sense out of a range of phenomena. Research has shown that
people use various learning styles to perceive and process
information. For these reasons, we structured the scientific content
in multiple ways at our website, to engage a range of learners in a
variety of meaningful ways.

Novices and children often prefer interactive learning experiences,
which can motivate them to stay engaged until they achieve some
payoff. The interactive modules and experiments of the black hole
website are tailored for this audience. They offer goal-based
scenarios, to motivate learning and a sense of accomplishment when
completed. By contrast, experts and adults often prefer formal
organization and ready access to information that they already know
about and just want to locate. To reach this audience, the site also
contains an encyclopedia of FAQs --- frequently asked questions ---
about the physics and astronomy of black holes, as well as a detailed
glossary.

We are currently working on a modified version of the web site that
can be used as a kiosk exhibit in museums, science centers and
planetaria. This will further broaden the audience of the
project to include those people who do not have access to the Internet,
or who typically do not use it to broaden their knowledge.

\section{Current Science}
\label{s:science}

The entire site is based on the most recent scientific knowledge.
Recent results and ongoing projects are highlighted where possible.
We will continue to update the site as new discoveries
emerge. Astronomical images from state-of-the-art telescopes are used
in abundance, with a particular focus on Hubble. Extensive links are
provided to other websites on the Internet, which allows users to
explore specific subjects in more detail.

Even with its sharp focus on black holes, our site strives to teach
much more. We illustrate how humans can understand the universe by
detailed observations of the night sky. We teach basic concepts, like
light and gravity. We show how different perspectives can enrich our
understanding of nature. We illustrate the many wonders of our
universe, and demonstrate its scale by alerting the user to what is
really near and what is really far away. And we highlight the many
things about black holes --- and our universe --- that we still do not
understand.

In the broadest sense, our goal has been to show that even the most
mysterious of things can be understood with the combined application
of human thinking and powerful technology. We hope to convey the
importance of scientific thought and to instill an appreciation for
learning and an interest in science, especially in the younger
generation of users.


\acknowledgments We are grateful to the many people who assisted with the 
creation of the website, or who provided graphics, animations, or
software. A full list of credits is available at\\
http:/$\!$/hubblesite.org/discoveries/black\_holes/credits.html . We
would also like to thank Bob Brown for editorial assistance with this
article.

\clearpage




\ifsubmode\else
\baselineskip=10pt
\fi








\ifsubmode\else
\baselineskip=14pt
\fi


\newcommand{\figcapone}{At the first module, a viewfinder can be 
moved across the sky to find and collect night sky objects. The
objects can be viewed at different wavelengths and a text box
indicates whether or not a black hole has been found. The "Learn More"
button provides access to a wealth of background
information.\label{f:one}}

\newcommand{\figcaptwo}{At the second module, the user undertakes a trip in an
animated starship to a nearby black hole, encountering many intriguing
objects along the way. The cockpit dials show the current speed and
distance traveled. Here, the traveler has arrived at Cygnus X-1 and
witnesses how a stellar-mass black hole accretes mass from its binary
companion star.\label{f:two}}


\ifsubmode
\figcaption{\figcapone}
\figcaption{\figcaptwo}

\clearpage
\else\printfigtrue\fi

\ifprintfig


\begin{figure}
\epsfxsize=0.8\hsize
\centerline{{\tt figure provided as jpeg}}
\ifsubmode
\vskip3.0truecm
\setcounter{figure}{0}
\addtocounter{figure}{1}
\centerline{Figure~\thefigure (part 1)}
\else
\figcaption{\figcapone}
\fi
\end{figure}


\begin{figure}
\epsfxsize=0.8\hsize
\centerline{{\tt figure provided as jpeg}}
\ifsubmode
\vskip3.0truecm
\addtocounter{figure}{1}
\centerline{Figure~\thefigure (part 1)}
\else
\figcaption{\figcaptwo}
\fi
\end{figure}


\fi


\end{document}